# Critical dynamics and domain motion from permittivity of the electronic ferroelectric (TMTTF)$_2$AsF$_6$

Serguei Brazovskii[a,b*], Pierre Monceau[c], Felix Ya. Nad[c]

[a] *LPTMS, UMR 8626, CNRS & University Paris-Sud, Bat. 100, Orsay, F-91405 France*
[b] *International Institute of Physics, 59078-400 Natal, Rio Grande do Norte, Brazil*
[c] *CNRS & University Grenoble Alpes, Inst. NEEL, F-38042 Grenoble, France*

**Abstract**

The quasi one-dimensional organic conductor (TMTTF)$_2$AsF$_6$ shows the charge ordering transition at $T_{CO}$=101K to a state of the ferroelectric Mott insulator which is still well conducting. We present and interpret the experimental data on the gigantic dielectric response in the vicinity of $T_{CO}$, concentrating on the frequency dependence of the inverse 1/ε of the complex permittivity ε=ε′+iε″. Surprisingly for a ferroelectric, we could closely approach the 2$^{nd}$ order phase transition and to deeply reach the critical dynamics of the polarization. We could analyse the critical slowing-down when approaching $T_{CO}$ from both sides and to extract the anomalous power law for the frequency dependence of the order parameter viscosity. Moreover, below $T_{CO}$ we could extract a sharp absorption feature coming from a motion of domain walls which shows up at a frequency well below the relaxation rate.

## 1. Electronic ferroelectricity in (TMTTF)$_2$X

The importance of electron correlation effects resulting from long-range Coulomb interactions in organic quasi one-dimensional compounds (TMTTF)$_2$X has been revealed in dielectric measurements by the discovery [1] of the ferroelectric (FE) ground state, which appears in the temperature range (40–200K), well above the temperatures of conventional magnetic (SDW) or (spin)-Peierls ordering (see reviews [2,3]; below we shall refer only to later or unaccounted there publications). Microscopically, this is the charge ordered (CO) state characterized by the spontaneous charge disproportionation among neighboring TMTTF molecules and by the concomitant homogeneous displacements of the counterions X, see [4].

Even in the background - well away from the transition temperature $T_{CO}$, the dielectric constant reaches a big value ε′~10$^3$ which is at least two orders of magnitude higher than estimations of a contribution from the ionic displacements. The enhancement factor has been derived [3] as ~(ℏω$_p$/Δ)$^2$ where ω$_p$ is the plasma frequency and Δ is the gap coming from the CO. That allows to invoke a more recent [5] nick-name: the "electronic ferroelectric", and to put in conjunctions the observations in (TMTTF)$_2$X with the ferroelectricity in more insulating donor-acceptor (DA) crystals showing the neutral-ionic phase transition [5-7] via the intersite D-A charge transfer concomitant with the lattice dimerization.

---

[*] Corresponding author, e-mail: brazov@lptms.u-psud.fr

Moreover, unexpectedly for ferroelectrics with their tendency to interrupt the way to the 2nd order transitions by 1st order jumps, in (TMTTF)$_2$X the transition can be approached so closely that ε′ reaches an astonishing value exceeding 10$^6$. This is more than three orders of magnitude higher than the peak value in other organic ferroelectrics of a similar "electronic" nature [5-7], even if their maximal permittivity ~1000 was already called "gigantic" in comparison with traditional ferroelectrics.

The common backgrounds and distinction with respect to more conventional ferroelectrics materials is not only a quantitative dominance of electronic polarization over the ionic one. This is also that the transition is symmetrically predefined and microscopically understandable on basis of electronic correlations. This principle is generalizeable to a predictable design of other electronic ferroelectrics, including the conjugated polymers [8,9] which can open a way to true applications.

The microscopic picture of ferroelectricity in organic crystals is based on two coexisting symmetry lowering effects [3]: the dimerization of bonds (which is build-in) and the dimerization of sites (which comes as a spontaneous symmetry breaking of the CO) – the roles are reversed with respect to DA systems [5-7]. The interference of dimerizations lifts the inversion symmetry opening the way for the FE polarization. Moreover, each of the dimerizations contributes to the band folding which makes $4k_F$ to be the new reciprocal period resulting in that the new band becomes effectively half filled. That leads to the state of the Mott insulator which is beautifully confirmed by a perfect spin-charge separation, see [3]. This state can be viewed and described as a Wigner crystal of electrons or equivalently as the $4k_F$ (see [4]) CDW, both displaced off-centre. The driving force behind the transition is the energy gained while falling from the parent metal to the Mott insulator [3].

The breaking of the space inversion symmetry in the CO-FE phase in another, two-dimensional organic compounds α-BEDT was confirmed [10] by optical second-harmonic generation (SHG) which appears only below $T_{CO}$. Ferroelectric domains were also directly visualized by the SHG microscopy [10]. Time resolved pump-probe experiments [11] show that the SHG signal is suppressed by photon-pumping and is recovered in the time scale of picosecond. This fast photo-response indicates that the essential of the electric polarization is not due to lattice distortion but to the electronic charge order. SHG signal has been measured [12] also in several Fabre salts (X=SbF$_6$,ASF$_6$,PF$_6$) below the respective temperatures 156K, 102K and 65K just at which the dielectric permittivity diverges, yielding the definite proof of the ferroelectricity in the new phase.

## 2. Dielectric permittivity

The data we used here are those determined from low frequency permittivity measurements in (TMTTF)$_2$AsF$_6$ as published in Refs. [2,13]. What is new is their analysis and results from that. Among many studied compositions we shall use data for the very homogeneous (TMTTF)$_2$AsF$_6$ samples which shows sharpest divergent anomalies in temperature dependences near $T_{co}$=101K. Recall that the perfect Curie-Weise-Landau anomaly ε′(T) ~ 1/|T-T$_{co}$| spreads widely over 30K on both sides of $T_{co}$, see Fig.1a. So our choice of two representative temperatures T≈ T$_{co}$ ±4K puts us well within the region of critical fluctuations.

The real part of the dielectric permittivity ε′(T,f) in a broad range of T and f is shown in Fig.1a. Above $T_{co}$, the frequency dependence of the imaginary part of the permittivity ε″(f) exhibited a maximum at $f_m$, with a nearly symmetrical form both above and below $f_m$ – Fig.1b. On the contrary, already for a small temperature decrease below $T_{co}$, the ε″(f) curve develops an additional low-frequency shoulder [2,13]. Interpreting $f_m$ as coming from some effective relaxation time τ = 1/(2π$f_m$), the τ(T) dependence shows [2,13] a growth to a thermal-activated behaviour, τ~exp(Δ/$k_B$T) with Δ/$k_B$=310K (Fig.3b) in excellent agreement with the activation energy 315K deduced from conductivity measurements below $T_{co}$ [14]. Notice the high quality of the Arhenius plot for **τ**, which ensures the precise determination of its activation energy. At low T the damping of the FE polarization comes from normal carriers - actually they are solitons as excitations of the 1D Mott insulator [3].

In a reduced temperature interval $T_{co}$±10K, the relaxation time diverges which qualitatively corresponds to the Landau-Khalatnikov (LK) theory [15] of a classical ferroelectric transition for which it is known that τ~1/|T-T$_{co}$|. This divergence corresponds to the softening of the oscillating mode responsible for the observed ferroelectric transition.

For a more detailed analysis, in the following we will use the inverse complex permittivity μ = 1/ε = μ′+i μ″ rather than the conventional complex ε = ε′+i ε″. In fact, to account for nonlinearities in the critical regime, μ is more fundamental than ε in the calculation of fluctuating ferroelectric polarization

P(r,t) as it intervenes in the energy functional F, such as in space-time (see also the eq. (2) below):

$$F = \frac{1}{2}\int dr P(r,t)\mu(r,t)P(r,t) \quad (1)$$

Moreover, for studies of critical relaxation, the use of $\mu \sim (1/\tau+i\omega)$ (as in the LK theory) is simpler and more straightforward than the commonly used information as in Fig.1b, from the peak in $\varepsilon'' \sim \omega / (1 + (\omega\tau)^2)$.

## 2.1. Above $T_{CO}$

The advantage of using $\mu$ is clearly demonstrated by comparison of Figs. 1 and 2 in which $\varepsilon''$ (Fig.1b) and $\mu'$ and $\mu''$ (Fig.2a) are plotted at the temperature T = 105K (T-$T_{CO}$=4K). The symmetrical shape of $\varepsilon''$ around its maximum is transformed, in the whole frequency range, into the linear log-log dependence for $\mu''$ which yields the power law $\mu'' \sim f^n$ with n = 0.86 ± 0.01. This truly unexpected unconventional power n≠1 directs us to the dynamical critical phenomena in deeply fluctuating regime [15,16].
The linear fit of Fig.2a covers symmetrically both slopes of each part around the maximum of $\varepsilon''$. In this scheme, $\mu'(f)$ may have only week f dependence (it is constant in the LK approximation) and it crosses $\mu''(f)$ at the frequency $f_m=1/\tau$, corresponding to the maximum in the $\varepsilon''(f)$ dependence. The strong f dependence of $\varepsilon'(f)$ seen in Fig.1b happens to be the consequence of the dependence $\mu''(f)$ which is disentangled from $\mu'(f)$ in the $\mu$-representation of data.

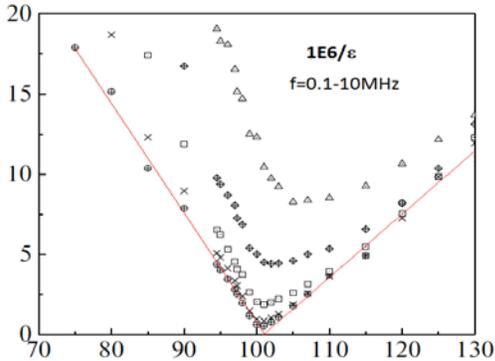

Fig.1a (data from Ref. [13]) : plots of $1/\varepsilon'(T,f)$ as functions of T at different f. Red lines are drawn to the eye to indicate the law $\varepsilon'(T) \sim 1/|T-T_{co}|$ with slopes ratio being just 2:1 as prescribed by the Landau theory.

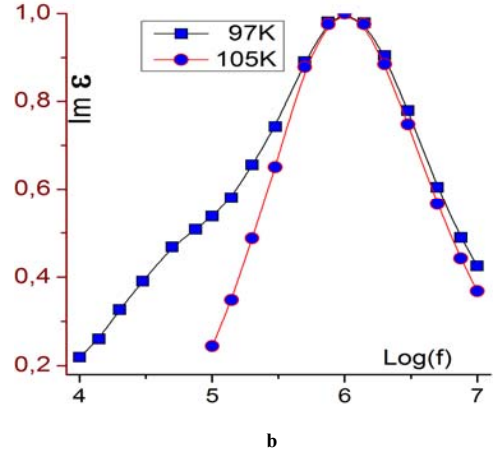

Fig.1b Lin-log plot of $\varepsilon''(f)$ above $T_{CO}$ at T= 105K (circles) and below $T_{CO}$ at T=97K (squares).

The difference between the experimental exponent 0.86 and the value 1 which would correspond to the mean field LK theory for critical relaxation allows to access the dynamical scaling at the criticality [15-17]. However, more detailed data are needed to approach nearer the phase transition to extract the scaling in both f and T-$T_c$.

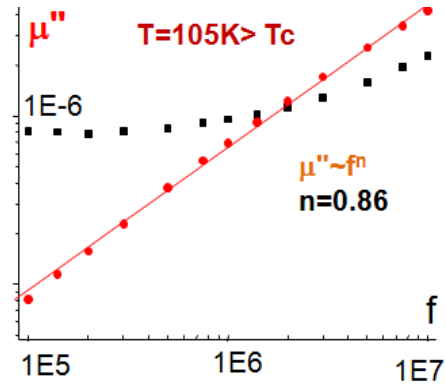

Fig.2a. Frequency dependences of $\mu'$ (black squares) and $\mu''$ (red circles) for the same data as in Fig.1 at T=105K=$T_{CO}$+4K.

## 2.2. Below $T_{CO}$

The frequency dependencies of $\mu'$ and $\mu''$ below $T_{CO}$ at a temperature $T_{CO}$-T = 4K are plotted in the log-log scale in Fig.2b. The power law for $\mu''(f)$ is achieved again, while at a shorter interval of higher frequencies which reflects the asymmetry of the $\varepsilon''(f)$ dependence (see Fig.3 in Refs. [2,13]). The critical exponent in $\mu'' \sim f^n$ is now n=0.78 which is close to but significantly different than its value above $T_{CO}$.

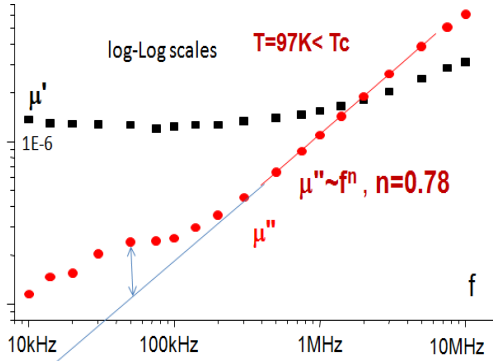

Fig.2b. Frequency dependences of μ′ (black squares) and μ″ (red circles) for the same data as in Fig.1 at T = 97K=$T_{CO}$- 4K. The blue line is the low f extrapolation of the power law dependence at higher frequency.

By extrapolation of the power law towards lower frequency (the blue line in Fig. 2b) and by subtracting this contribution, we can single out the contribution of the low frequency shoulder in μ″. In this way, we disentangle the low frequency feature from the main relaxation as plotted in Fig. 3, now in the linear scale.

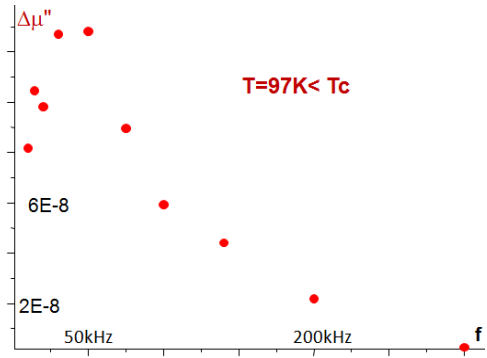

Fig. 3a: Low frequency increment of μ″ in linear scales. The peak is attributed to a motion of FE domains. Right (based on data from Ref. [13]): T dependence (the Arrhenius plot) of the relaxation time τ and the extraction (the straight red line drawn by eye) of the constant activation energy at low T.

The low frequency contribution appears as a resonance or a threshold around 50kHz - well below the relaxation rate - that can be attributed to slow oscillations of pinned ferroelectric domain walls.

More detailed data are needed first to approach nearer the phase transition for a full description of criticality and for study of dynamics of domain walls.

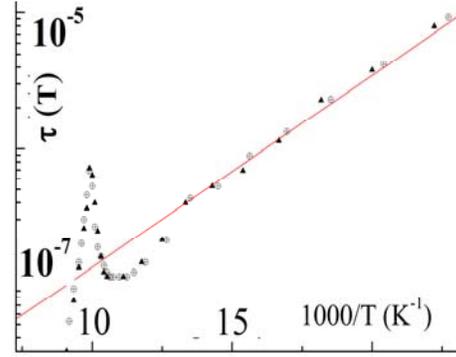

Fig. 3b: (based on data from Ref. [13]): T dependence (the Arrhenius plot) of the relaxation time τ and the extraction (the straight red line drawn by eye) of the constant activation energy at low T.

*2.3 Interpretation of the anomalous scaling.*

The sense of the nontrivial exponent *n* can be clarified if we write vaguely the operator μ in eq. (1) as

$$\mu = \langle \eta \partial/\partial t + \delta^2 H / \delta P^2 \rangle \quad (2)$$

where the average $\langle\rangle$ is taken over all fluctuations at lengths shorter than the macroscopic one. Our results show that the viscosity constant η becomes a retarded function rather than a bare constant.

The common hypothesis of the dynamical scaling (see [15-17]) usually does not distinguish the real and the imaginary parts coming from the second and the first terms in (2); they are mixed up in the common scaling of the average response function akin to ε in our case. The RG theories (see [16]) usually assume η=cnst for the conserved nondegenerate order parameter which is our case. The question is open: either we arrived to a general unattended feature of the dynamical scaling, or we are set in a particular limit when the thermodynamic fluctuations are still conventional (Landau theory) while the viscosity becomes anomalous [18], or there is something special in our case of the electronic ferroelectrics with its Coulomb interactions and the viscosity by friction to solitonic current carriers.

**3. Conclusion**

The ferroelectric state in $(TMTTF)_2AsF_6$ as well as in other Fabre salts coexists with a low-activated spinless conductivity which gives rise to the unusual - "ferroelectric Mott semiconductor" state. The intrinsic conductivity provides the screening of outgoing electric field which eliminates the equilibrium domain structure and the resulting hysteresis which are common to ferroelectrics. That seems to lead to complete repolarisation under the ac field, and even to unusual mono-domain state of the whole sample. The phase transition acquires a highly pronounced second order character which normally does not take place in ferroelectrics.

By the simultaneous analysis of the real part and the imaginary part of the inverse complex dielectric permittivity in $(TMTTF)_2AsF_6$, we have been able to separate several dynamic regimes: a) the critical slowing down characterized by the anomalous power law with exponents 0.86 and 0.78 at temperature respectively 4K above and 4K below the ferroelectric transition;
b) the low frequency response associated to sweeping or oscillating domain walls below $T_{CO}$.
More experiments are necessary at temperatures even much closer to $T_{CO}$ as well as for larger frequency range at low T.

**Acknowledgements.**
S.B. and P.M. devote this article to their deceased co-author Felix Ya. Nad.